\documentclass[aps,prb,twocolumn,superscriptaddress,showpacs]{revtex4-1}
\usepackage[dvips]{graphicx}

\begin{document}

\title{Doping dependence of the spin excitations in Fe-based superconductors Fe$_{1+y}$Te$_{1-x}$Se$_{x}$}

\author{A. D. Christianson}
\author{M. D. Lumsden}
\author{K. Marty}
\author{C. H. Wang}
\author{S. Calder}
\author{D. L. Abernathy}
\author{M. B. Stone}
\author{H. A. Mook}
\author{M. A. McGuire}
\author{A. S. Sefat}
\author{B. C. Sales}
\affiliation{Oak Ridge National Laboratory, Oak Ridge, Tennessee 37831, USA}
\author{D. Mandrus}
\affiliation{Oak Ridge National Laboratory, Oak Ridge, Tennessee 37831, USA}
\affiliation{Department of Materials Science and Engineering, University of Tennessee, Knoxville, Tennessee 37996, USA}

\author{E. A. Goremychkin}
\affiliation{School of Physics and Astronomy, University of Southampton, Southampton SO17 1BJ, UK}
\affiliation{ISIS Facility, Rutherford Appleton Laboratory, Chilton, Didcot, Oxon OX11 OQX, United Kingdom}

\date{\today}

\begin{abstract}
The Fe$_{1+y}$Te$_{1-x}$Se$_{x}$ series of materials is one of the prototype families of Fe-based superconductors.  To provide further insight into these materials we present systematic inelastic neutron scattering measurements of the low energy spin excitations for x=0.27, 0.36, 0.40, 0.49.  These measurements show an evolution of incommensurate spin excitations towards the (1/2 1/2 0) wave vector with doping.  Concentrations (x=0.40 and 0.49) which exhibit the most robust superconducting properties have spin excitations closest to (1/2 1/2 0) and also exhibit a strong spin resonance in the spin excitation spectrum below T$_c$. The resonance signal appears to be closer to (1/2 1/2 0) than the underlying spin excitations.  We discuss the possible relationship between superconductivity and spin excitations at the (1/2 1/2 0) wave vector and the role that interstitial Fe may play.

\end{abstract}



\maketitle

\section{Introduction}

The link between superconductivity and magnetism in many unconventional superconductors has been a recurring avenue of investigation.  The heavy fermions \cite{stewart2001}, the cuprates \cite{lee2006}, and the Fe-based \cite{lumsden2010a} superconductors all exhibit some form of magnetism across vast regions of their phase diagrams.  The discovery of superconductivity in LaFeAsO$_{1-x}$F$_x$ \cite{kamihara2008} and numerous related materials has presented an opportunity to study the relationship of superconductivity and magnetism in a new context.

The Fe$_{1+y}$Te$_{1-x}$Se$_{x}$ family of superconductors holds promise as a model system for the investigation of unconventional superconductivity in the Fe-based materials.  This is in large part due to the apparent chemical and structural (tetragonal space group $P4/nmm$) simplicity, but also due to the ease at which the physical properties can be tuned. FeSe was found to be superconducting at 8 K\cite{hsu2008}.  Subsequently Fe stoichiometry\cite{mcqueen2009} was found to be critical and that both pressure \cite{mizuguchi2008} and chemical substitution\cite{yeh2008} can be employed to tune the superconducting properties.  In particular, substitution of tellurium for selenium increases T$_c$ to 15 K for x$\sim$0.5 with additional doping leading to the destruction of the superconducting state for x$\sim$0.3-0.4 and the formation of a magnetically ordered state\cite{khasanov2009,liu2010}.

The long range magnetically ordered state in Fe$_{1+y}$Te is characterized by a sizable magnetic moment of $\sim$ 2 $\mu_B$ and a  propagation vector of ($\delta$ 0 0.5) where $\delta$ $\sim$ 0.5 with the actual value exhibiting a strong dependence on Fe stoichiometry\cite{bao2009,li2009}.  With Se doping the magnetic order becomes short range and incommensurate before disappearing for x$\sim$0.15\cite{khasanov2009, bao2009, katayama2010}.  In the same region of the phase diagram where the (0.5 0 0.5) type magnetic order disappears two-dimensional incommensurate spin fluctuations develop near (0.5 0.5 0) or in square lattice notation ($\pi$ 0) as is typically found in other Fe-based superconductors such as in optimally doped Ba(Fe$_{1-x}$Co$_x$)$_2$As$_2$ \cite{lumsden2009}.

\begin{table}[htp]
\caption{\label{sample} Summary of samples and experimental conditions.  Sample stoichiometry has been determined by energy dispersive x-ray scattering measurements.  T$_c$ is determined from magnetic susceptibility measurements on small pieces ($\sim$10 mg) cut from the larger crystals.  FS denotes filamentary superconductivity as concluded from magnetic susceptibility measurements.}
\begin{ruledtabular}
\begin{tabular}[b]{ccccc}
 sample    &  mass (g) &  T$_c$ (K) & instruments \\
\hline
Fe$_{1.04}$Te$_{0.73}$Se$_{0.27}$ &16.9&  FS & HB1\\
Fe$_{1.04}$Te$_{0.64}$Se$_{0.36}$ &16.2&  FS & HB3\\
Fe$_{1.02}$Te$_{0.60}$Se$_{0.40}$ &  15.0 & 13 &HB3\\
Fe$_{1.00}$Te$_{0.51}$Se$_{0.49}$ &  15.5 & 14 &ARCS, HB3\\
\end{tabular}
\vspace{-2mm}
\end{ruledtabular}
\end{table}

The magnetic excitations in the Fe$_{1+y}$Te$_{1-x}$Se$_{x}$ family have been the subject of considerable experimental scrutiny \cite{qiu2009,lumsden2010b,mook2010,xu2010,wen2010,katayama2010, li2010,tsyrulin2012, stock2012, babkevich2010, argyriou2010, lee2010, marty2012,thampy2012}. Generally, the spin excitations in the region of the phase diagram where superconductivity is found $ 0.4 \leq x \leq 1 $ originate at incommensurate wave vectors near (1/2 1/2) and extend to energies on the order of 300 meV.  At lower energies the onset of superconductivity is found to significantly alter the spin excitation spectrum.  At T$_c$ a gap in the spin excitation spectrum opens with an associated transfer of spectral weight to an energy of $\sim$2$\Delta$ where $\Delta$ is the superconducting gap. This extra intensity is referred to as the spin resonance and is commonly interpreted as indicating a sign change of the superconducting order parameter on different portions of the fermi surface and thus provides evidence for the unconventional superconductivity found in the Fe-based superconductors.  In the Fe$_{1+y}$Te$_{1-x}$Se$_{x}$ family and indeed many other Fe-based superconductors the resonance occurs at an energy of $\sim$ 5 k$_B$T$_c$ \cite{lumsden2010a,insov2011}.     Thus the low energy spin excitation spectrum provides a window into the interplay of magnetism and superconductivity.  Previously, we\cite{lumsden2010b,mook2010} and others\cite{xu2010,babkevich2010}, have found evidence that spin exciations near (1/2 1/2) are associated with the most favorable superconducting properties in the Fe$_{1+y}$Te$_{1-x}$Se$_{x}$ series.  Here in an attempt to probe the relationship between magnetism and superconductivity in greater detail, we have performed a systematic study of the spin excitations in several concentrations, x=0.27, 0.36, 0.40, 0.49, synthesized via the same method. We find an evolution of the spin excitation spectrum towards the (1/2 1/2) wave vector with doping and in particular those samples which exhibit the strongest superconducting properties have spin excitations closest to (1/2 1/2).  Moreover, in superconducting samples the intensity in the resonance is closer to (1/2 1/2) than the normal state spin excitations.  Related work has been previously published in the following papers \cite{lumsden2010b,mook2010,marty2012}.

\begin{figure}[h]
\begin{center}
		\includegraphics[width=0.45\textwidth]{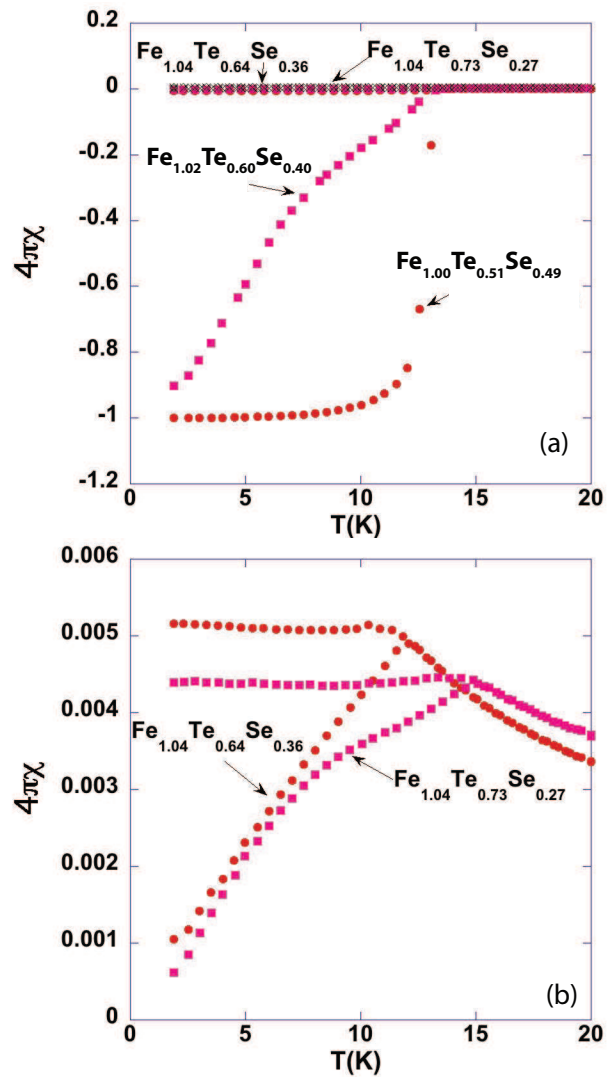}
		\caption{ (Color online) Magnetic susceptibility data for Fe$_{1+y}$Te$_{1-x}$Se$_{x}$.  The measured samples were cut from the crystals used in the neutron scattering experiments.  For x=0.40 and 0.49 bulk superconductivity is observed whereas for x=0.36 and 0.27 only small hints of a diamagnetic contribution to the magnetic susceptibility are observed.   Field cooled and zero field cooled data were collected in all cases with an applied field of 20 oersted.  }
		\label{susceptibility}
	\end{center}
\end{figure}

\begin{figure*}[t]
\begin{center}
		\includegraphics[width=\textwidth]{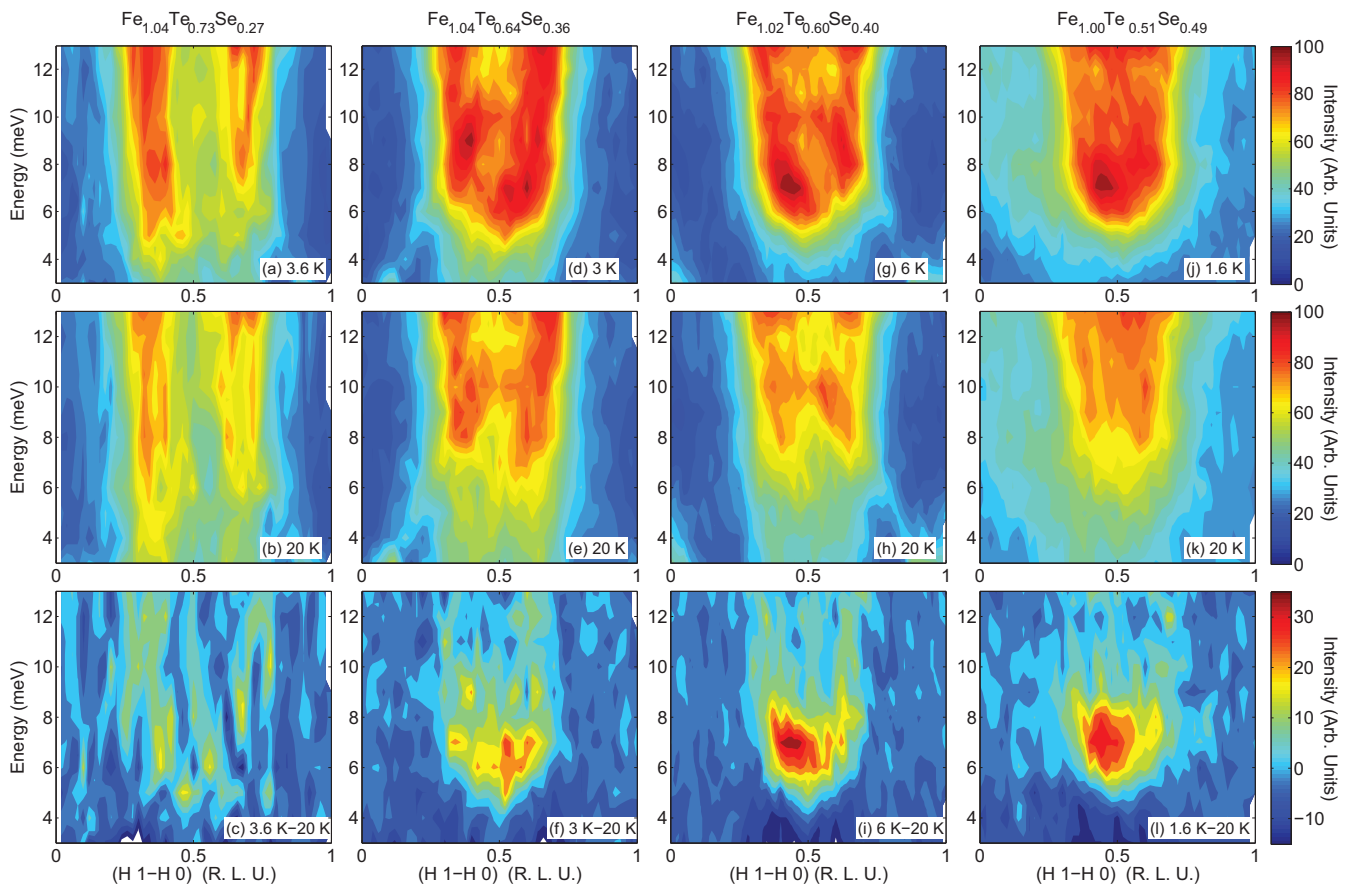}
		\caption{ (Color online) Comparison of inelastic neutron scattering data along the (H 1-H 0) direction as a function of energy for different concentrations x.  All data was collected with triple-axis instruments as indicated in table \ref{sample}.  (a)-(c) x = 0.27, (d)-(f) x = 0.36, (g)-(i) x = 0.40, (j)-(l) x = 0.49.  The top row is the low temperature data. The second row is data collected at 20 K.  The third row of data is the result of subtracting the 20 K data from the base temperature data. Each column of data has been normalized consistently, however care should be taken when comparing the data between columns as data were collected with different instruments and under different experimental conditions with samples of differing mass.}
		\label{mega}
	\end{center}
\end{figure*}

\begin{figure}[h]

\begin{center}
		\includegraphics[width=0.45\textwidth]{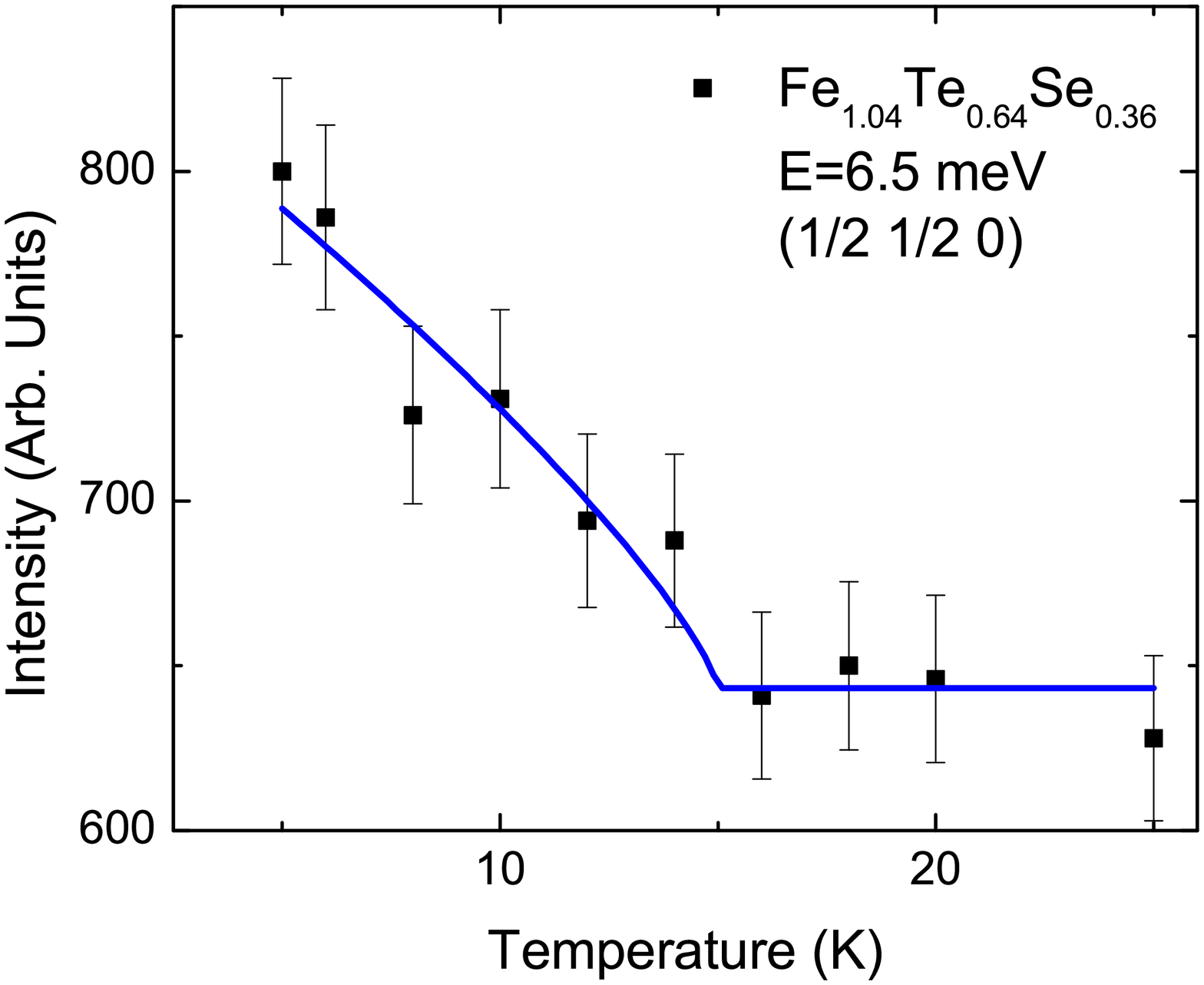}
		\caption{ (Color online) Temperature dependence of the inelastic neutron scattering intensity for Fe$_{1.04}$Te$_{0.64}$Se$_{0.36}$ at 6.5 meV at (1/2 1/2).  The line is a guide to the eye.}
		\label{65_35}
	\end{center}
\end{figure}

\section{Experimental Details}

All samples were synthesized by a modified Bridgeman method as described in Ref. \onlinecite{sales2009}.  The sample stoichiometry was determined with energy dispersive x-ray scattering measurements.  Within the resolution of these measurements no sample inhomogeneity was detected.  The magnetic susceptibility was measured with a Quantum Design MPMS with an applied field of 20 oersted.   The resulting sample compositions, masses and T$_c$s and neutron scattering instruments where the samples were measured are given in table \ref{sample}. Room temperature lattice parameters have been determined from powder x-ray diffraction on samples grown as described in Ref. \onlinecite{sales2009}.  The c-axis lattice parameters, which are the most sensitive to doping level, are 6.163 $\AA$ (x=0.27), 6.109 $\AA$ (x=0.36), 6.060 $\AA$ (x=0.40), and 6.034 $\AA$ (x=0.49).   No structural changes were observed in the samples studied here, although the resolution of the experimental configuration utilized in the neutron scattering measurements was low and subtle structural changes such as the small monoclinic or orthorhombic distortions reported in the literature would be unobservable\cite{martinelli2010,gresty2009}.

Figure \ref{susceptibility} shows the magnetic susceptibility data for small pieces ($\sim$ 10 mg) cut from the large neutron scattering crystals.  Both zero field cooled and field cooled data were collected.  In the units of Fig. \ref{susceptibility} a value of -1 indicates complete screening.   Figure \ref{susceptibility}(a) shows that both x=0.40 and x = 0.49 exhibit bulk superconductivity, where as x=0.36 and x=0.27 do not.  However, on the magnified scale in Fig. \ref{susceptibility}(b) there does appear to be a reduction of the susceptibility that may be due to a small superconducting volume fraction.  The susceptibility data would then suggest that the x=0.40 and x=0.49 are most strongly superconducting while the volume fraction in the x=0.27 and 0.36 samples is small. As will be discussed later small pieces cut from the larger samples may not be entirely representative.

Several inelastic neutron scattering instruments were used in this study.  The HB1 and HB3 thermal triple-axis spectrometers were used at the High Flux Isotope Reactor (HFIR) at Oak Ridge National Laboratory.  Collimations of 48'-60'-60'-240' (x=0.27) (HB1), 48'-60'-80'-120' (x=0.36) (HB3), 48'-40'-40'-120' (x=0.49) (HB3), and 48'-60'-80'-120' (x=0.4) (HB3) were used.  For both instruments pyrolitic graphite monochromators and analyzers were used.  The final energy was fixed at 14.7 meV for both HB1 and HB3.  Pyrolitic graphite filters were used to suppress higher order contamination on both instruments. Time-of-flight inelastic neutron scattering measurements were performed on the ARCS direct geometry chopper spectrometer at the Spallation Neutron Source at Oak Ridge National Laboratory.  An incident energy of 25 meV was used and measurements were collected in angular steps of one degree with the HHL scattering plane horizontal at 4 K (below Tc) and 25 K (above Tc) for FeTe$_{0.51}$Se$_{0.49}$.

\section{Results and Discussion}

Figure \ref{mega} shows a comparison of the scattering along (H 1-H 0) as a function of energy transfer for x=0.27, 0.36, 0.40, and 0.49.  The top row ((a),(d),(g), (j)) displays the data collected at low temperatures (below T$_c$ for the superconducting samples). The middle row ((b), (e), (h), (k)) shows the data collected at 20 K. The bottom row shows the result of subtracting the data collected at 20 K from that collected at low temperature.  As has been shown previously \cite{lumsden2010a} the contrast between the x=0.27 data and the x=0.49 data is readily apparent.   The inelastic spectral weight in the superconducting sample (x=0.49) is much closer to (1/2 1/2) than in the nonsuperconducting sample (x=0.27).  A positive signal in the subtracted data is indicative of the well studied spin resonance \cite{bao2009,mook2010,argyriou2010, lee2010}.   There is almost a negligible, broadly distributed signal in this regard in the nonsuperconducting sample (x=0.27) while there is a strong, localized resonance in the superconducting sample (x=0.49).

The data for the other two concentrations, x=0.36 and x=0.40, are shown in the middle two columns and provides further information concerning the doping dependence of the spin excitations and the resonance. The magnetic susceptibility measurements presented in Fig. \ref{susceptibility} suggest that the x=0.36 sample does not exhibit bulk superconductivity.  Despite the apparent lack of superconductivity in the x=0.36 sample,  a well defined resonance is observed.  On the other hand, the intensity of the resonance in the x=0.40 sample is as strong as that found in the x=0.49 sample.   The subtracted data displayed in Fig. \ref{mega} (c), (f), (i), and (l) also shows that aside from the resonance and associated gapped spectral weight at low energies the spin wave velocity does not change significantly above and below T$_c$.

The temperature dependence of the scattering in the x=0.36 sample was examined to provide further insight into the appearance of the resonance like feature in a sample which is nominally not superconducting.  The inelastic neutron scattering intensity as a function of temperature at the (1/2 1/2) wave vector at 6.5 meV is shown in Fig. \ref{65_35}.   The data suggests a T$_c$ between 12 and 15 K (Fig. \ref{65_35}).  This temperature dependence is similar to the temperature dependence of the resonance for other members of the  Fe$_{1+y}$Te$_{1-x}$Se$_{x}$ family. This raises the question: why is a larger diamagnetic contribution not seen in the magnetic susceptibility?  The most obvious explanation is that a small piece cut from the edge of a larger sample is not entirely representative of the bulk.  However, the weak diamagnetic signal is consistently observed from multiple sample batches grown with the method in Ref. \onlinecite{sales2009}.  An additional possibility is that sample inhomogeneity \cite{zhu2013} results in inclusions of a superconducting phase such as x$\sim$0.5.  However, energy dispersive x-ray scattering measurements were not able to detect any chemical inhomogeneity.  A third possible explanation is that the interstitial Fe present in this concentration contributes a large Q=0 paramagnetic susceptibility which renders the full diamagnetic susceptibility impossible to observe.  Inelastic neutron scattering at the wave vector (1/2 1/2) would be relatively unaffected by the paramagnetic contribution to the susceptibility and thus in this case inelastic neutron scattering is probably more diagnostic of the superconducting volume fraction of the sample.

\begin{figure}[h]
\begin{center}
		\includegraphics[width=0.45\textwidth]{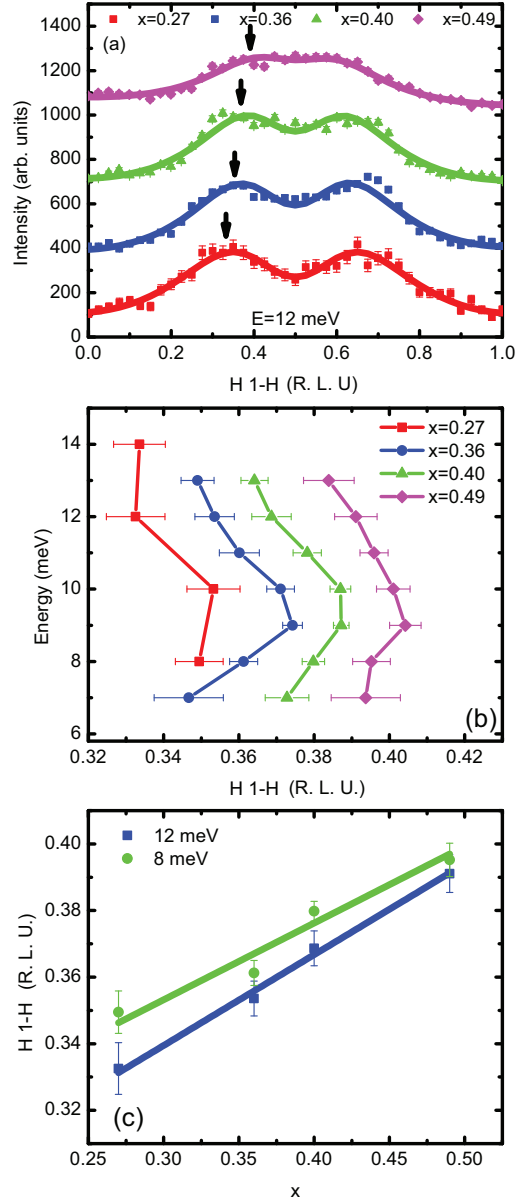}
		\caption{ (Color online) The dispersion of the spin excitations in Fe$_{1+y}$Te$_{1-x}$Se$_{x}$ at 20 K.  (a) A comparison of the q-scans at 12 meV for each concentration. The lines are fits with the Sato-Maki function as described in the text.   (b) Shows the dispersion extracted as explained in the text for x=0.27,0.36, 0.40, 0.49.(c) Shows the value of the h for each concentration at 12 and 8 meV.  The lines are guides to the eye.}
		\label{dispersion}
	\end{center}
\end{figure}

\begin{figure*}[t]
\begin{center}
		\includegraphics[width=\textwidth]{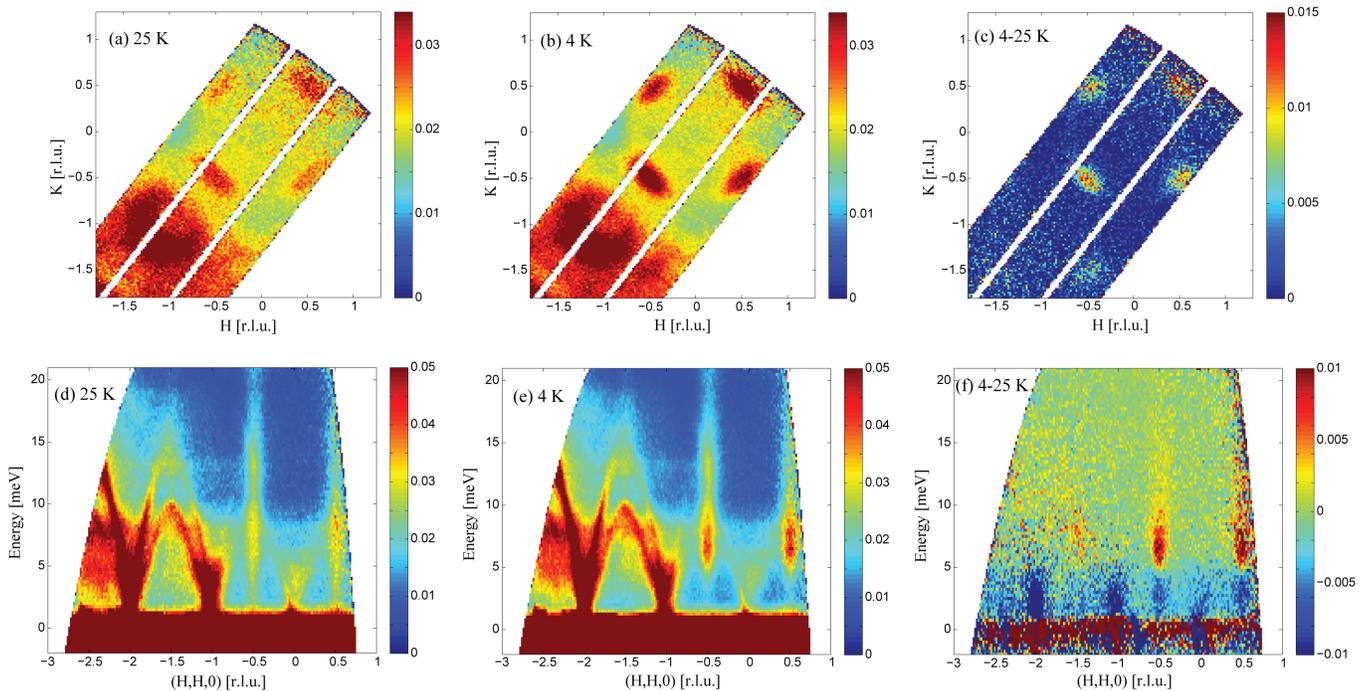}
		\caption{ (Color online) The spin excitations of the x=0.49 single crystal collected with ARCS. (a), (b) and (c) display the data in the (H K 0) plane at 6.5$\pm$1 meV for L=$\pm$2 r.l.u. for temperatures of 25 K, 4 K and the difference 4-25 K respectively. (d), (e) and (f) display the data in (H H 0) - Energy plane at (K, -K)=$\pm$0.1 r.l.u. and L=$\pm$2 r.l.u. for temperatures of 4 K, 25 K and the difference 4-25 K respectively}
		\label{arcs}
	\end{center}
\end{figure*}

A visual inspection of the data in Fig. \ref{mega} suggests that the evolution of the spin excitations towards (1/2 1/2) is not abrupt but proceeds gradually across the doping range studied here. To examine this in more detail we have fit the data with the Sato-Maki function which has been used previously to parameterize the spin excitations in the Fe$_{1+y}$Te$_{1-x}$Se$_{x}$ family as well as other materials \cite{lumsden2010a,satomaki1974, vignolle2007}.  The data and fits at 12 meV are shown in Fig. \ref{dispersion} (a). The data were collected at a temperature of 20 K which exceeds T$_c$ in all samples to avoid the influence of the signal from the resonance  The data shows a clear evolution towards (1/2 1/2) with increasing selenium concentration.  This is also the region of the phase diagram where T$_c$ is highest.
Figure \ref{dispersion}(b) shows the energy dependence of the scattering as a function of H along the (H 1-H 0) direction.  The points were extracted from fits of the inelastic neutron scattering data with the Sato-Maki function. There is an additional shift towards (1/2 1/2) in the 8-10 meV range for all concentrations.  This appears to be the case whether we use the Sato-Maki function to extract the dispersion, fit with Lorentzians, or simply take the peak positions for the concentrations where the peaks are well defined.  The energy range discussed here is above the energy range where others have reported an hourglass dispersion \cite{li2010,tsyrulin2012}. The concentration dependence of the position of the spin excitation is shown for 8 and 12 meV in Fig. \ref{dispersion} (c).   This suggests the evolution of the spin fluctuations toward the (1/2 1/2) position proceeds smoothly as a function of doping.

To clarify the origin of the shift of the spin excitations toward (1/2 1/2) near 9 mev, data on the x=0.49 single crystal were collected using the ARCS chopper spectrometer.  These data allow for simultaneous measurements of both the magnetic and lattice excitation spectrum and representative projections onto the (H,H,0) - Energy plane are shown in Fig. \ref{arcs} (d) and (e) for temperatures of 25 K and 4 K respectively.  The data clearly show the magnetic excitations located at the (1/2 1/2) and (-1/2 -1/2) wave vectors and also shows phonons in higher Brillouin zones.  The transverse acoustic (TA) phonon modes reach a zone boundary energy of about 9 meV while the longitudinal acoustic modes extend to about 14 meV.  The intensity of these modes clearly gets larger as the wave vector increases as expected for phonon scattering.  Despite the much weaker intensity near the (1/2 1/2) wave vectors, there is still phonon scattering present and the intersection of the TA modes and the magnetic excitations will produce enhanced spectral weight near 9 meV.  The influence of this phonon scattering will impact the Sato-Maki fits described above and provides an explanation for the shift observed in the 8-10 meV energy range.

To more definitively show that the resonant part of the spin excitation spectrum is closer to (1/2 1/2), ARCS data is presented in Fig. \ref{arcs}.  Panels (a) and (b) display data at 6.5$\pm$1 meV projected onto the (HK0) plane for x=0.49 above and below T$_c$.  The spectral weight due to the resonance is given by the difference between the data above and below T$_c$ and is shown in Fig. \ref{arcs} (c).  The difference data is more localized near the (1/2 1/2) wave vector than either the data above or below T$_c$ indicating that the resonance is considerably more isotropic than the underlying spin excitations in agreement with the findings for x=0.49 in Ref. \onlinecite{mook2010}.  Despite being more isotropic than the underlying spin excitations, the resonance is still elongated along the (H 1-H) direction in comparison to the (H H) direction.  A similar anisotropy of the resonance has been observed in other Fe-based superconductors\cite{hli2010}.    In addition, we note that Fig. \ref{arcs} (f) shows the data at 25 K subtracted from the data at 4 K projected onto the (H,H,0) - Energy plane (i.e. panel (d) subtracted from panel (e)).  This demonstrates the spin resonance at ($\pm$1/2 $\pm$1/2 0) and also shows a much weaker resonance at (-1.5,-1.5,0) consistent with the magnetic nature of this scattering.  The data also shows that the resonance signature, while sharply peaked around 7 meV, extends to much higher energy transfers and positive response in this difference plot is observed for energy transfers up to at least 15 meV. The large extent in energy of the resonance may indicate a significant anisotropy of the superconducting gap.


The results presented here suggest that spin fluctuations near (1/2 1/2) coincide with the best superconducting properties whereas spin fluctuations emanating from incommensurate positions are unfavorable for superconductivity.  This contention is supported by systematic changes in the normal state spin excitations with doping and by changes in the spin excitation spectrum produced by the onset of superconductivity.   Precisely how spin excitations near (1/2 1/2) influence superconductivity remains a matter of debate.  Proposals for $S_\pm$ pairing symmetry rely on nesting between a hole pocket at the $\Gamma$ point and an electron pocket at the M point\cite{mazin2008,kuroki2008}.  Within this picture spin fluctuations near (1/2 1/2) would be more favorable for superconductivity.  An interesting analogy exists with the 115 family of heavy fermion superconductors.  In these materials a higher T$_c$ is associated with commensurate antiferromagnetic spin fluctuations at (1/2 1/2 1/2)\cite{stock2008}. On the other hand, the presence of interstitial Fe in the Fe$_{1+y}$Te$_{1-x}$Se$_x$ series complicates a simple interpretation of the data. In particular, studies of the spin excitations as a function of interstitial Fe concentration, \textit{y}, show that interstitial Fe is both pair breaking and tends to push the spin excitations farther from (1/2 1/2)\cite{tsyrulin2012,stock2012,thampy2012}. The destruction of superconductivity by interstitial Fe has also been implicated by scanning tunneling microscopy data which shows nonsuperconducting regions centered around extra atoms\cite{lin2013}. In fact, if the large changes in Te doping are neglected, a smaller amount of interstitial Fe in the samples studied here also coincides with a tendency of spin excitations towards (1/2 1/2).  However, Te doping raises T$_c$ from 8 K in FeSe to 15 K in FeTe$_{0.5}$Se$_{0.5}$\cite{yeh2008}. Thus both Te doping and interstitial Fe play an important role in the superconductivity of Fe$_{1+y}$Te$_{1-x}$Se$_x$.

\section{Conclusions}

We have performed an extensive study of the low energy spin excitations in the Fe$_{1+y}$Te$_{1-x}$Se$_x$ family of Fe-based superconductors.  The same synthesis process has been used for all samples facilitating comparison across materials with different composition. We find that the samples with the most robust superconducting properties have spin excitations which are closest to the (1/2 1/2) wave vector.  We are also able to explain the deviations in the excitation spectrum in the vicinity of 8 to 10 meV energy transfer as being due to the influence of lattice excitations in this portion of the spectrum. The spectral weight in the spin resonance is also found to be closer to (1/2 1/2) than the underlying spin excitations.

\begin{acknowledgments}
We acknowledge useful discussions with T. A. Maier.  Research at ORNL is sponsored by the Scientific User Facilities Division and the Materials Sciences and Engineering Division, Office of Basic Energy Sciences, U.S. DOE.
\end{acknowledgments}

\end{document}